# SWING-BY IN THREE DIMENSIONS: CLOSED FORM SOLUTIONS

**Antonio Fernando Bertachini de Almeida Prado**
Instituto Nacional de Pesquisas Espaciais - INPE - Brazil
São José dos Campos - SP - 12227-010 – Brazil
Peoples' Friendship University of Russia Named After Patrice Lumumba
(RUDN University), 6, Miklukho-Maklaya Str., 117198 Moscow, Russia

## ABSTRACT

The main goal of the present research is to make a summary of analytical equations that can be found to calculate a swing-by maneuver in the three-dimensional space. Analytical equations based in the patched conics approximation are showed and they allow to calculate the variation in velocity, angular momentum, energy and inclination of the spacecraft that is involved in this maneuver. Based on that, it is possible to obtain expressions for particular cases, like the planar and the polar maneuver. The most important properties of this maneuver can demonstrated using thos equations, like: in the case of planar maneuver the variation in inclination can be only 180°, 0º, and -180°; the variation in inclination is symmetric with respect to the out of plane angle; a passage by the poles changes only the y-component of the angular momentum, keeping the energy and the inclination of the trajectory unchanged. The results show several maneuvers.

## INTRODUCTION

The swing-by maneuver is a very popular technique used to decrease fuel expenditure in space missions. The literature shows several applications of the swing-by technique. Some of them can be found in Swenson[1], that studied a mission to Neptune using swing-bys to gain energy to accomplish the mission; Weinstein[2], that made a similar study for a mission to Pluto; Farquhar and Dunham[3], that formulated a mission to study the Earth's geomagnetic tail; Farquhar, Muhonen and Church[4], Efron, Yeomans, and Schanzle[5] and Muhonen, Davis, and Dunham[6], that planned the mission ISEE-3/ICE; Flandro[7], that made the first studies for the Voyager mission; Byrnes and D'Amario[8], that design a mission to flyby the comet Halley; D'Amario, Byrnes and Stanford[9,10] that studied multiple flyby for interplanetary missions; Marsh and Howell[11] and Dunham and Davis[12] that design missions with multiple lunar swing-bys; Prado and Broucke[13], that studied the effects of the atmosphere in a swing-by trajectory; Striepe, and Braun[14], that used a swing-by in Venus to reach Mars; Felipe and Prado[15], that studied numerically a swing-by in three dimensions, including the effects in the inclination; Prado[16], that considered the possibility of applying an impulse during the passage by the periapsis; Prado and Broucke[17], that classified trajectories making a swing-by with the Moon. The most usual approach to study this problem is to divide the problem in three phases dominated by the "two-body" celestial mechanics. Other models used to study this problem are the circular restricted three-body problem (like in Broucke[18], Broucke and Prado[19], and Prado[20]) and the elliptic restricted three-body problem (Prado[21]).

The goal of this paper is to develop analytical equations for the variations of velocity, energy, angular momentum and inclination for a spacecraft that passes close to a celestial body. This passage, called swing-by, is assumed to be performed around the secondary body of the system. Among the several sets of initial conditions that can be used to identify uniquely one swing-by trajectory, the following five variables are used: $V_p$, the velocity of the spacecraft at periapsis of the orbit around the secondary body; two angles ($\alpha$ and $\beta$), that specify the direction of the periapsis of the trajectory of the spacecraft around $M_2$ in a three-dimensional space; $r_p$ the distance from the spacecraft to the center of $M_2$ in the moment of the closest approach to $M_2$ (periapsis distance); $\gamma$, the angle between the velocity vector at periapsis and the intersection between the horizontal plane that passes by the periapsis and the plane perpendicular to the periapsis that holds $\vec{v}_p$.

Fig. 1 shows the sequence for this maneuver and some important variables. It is assumed that the system has three bodies: a primary ($M_1$) and a secondary ($M_2$) bodies with finite masses that are in circular orbits around their common center of mass and a third body with negligible mass (the spacecraft) that has its motion governed by the two other bodies. The spacecraft leaves the point A, passes by the point P (the periapsis of the trajectory of the spacecraft in its orbit around $M_2$) and goes to the point B. The points A and B are chosen in a such way that the influence of $M_2$ at those two points can be neglected and, consequently, the energy can be assumed to remain constant after B and before A (the system follows the two-body celestial mechanics). The initial conditions are clearly identified in Fig.1. The distance $r_p$ is not to scale, to make the figure easier to understand. The result of this maneuver is a change in velocity, energy, angular momentum and inclination in the Keplerian orbit of the spacecraft around the central body.

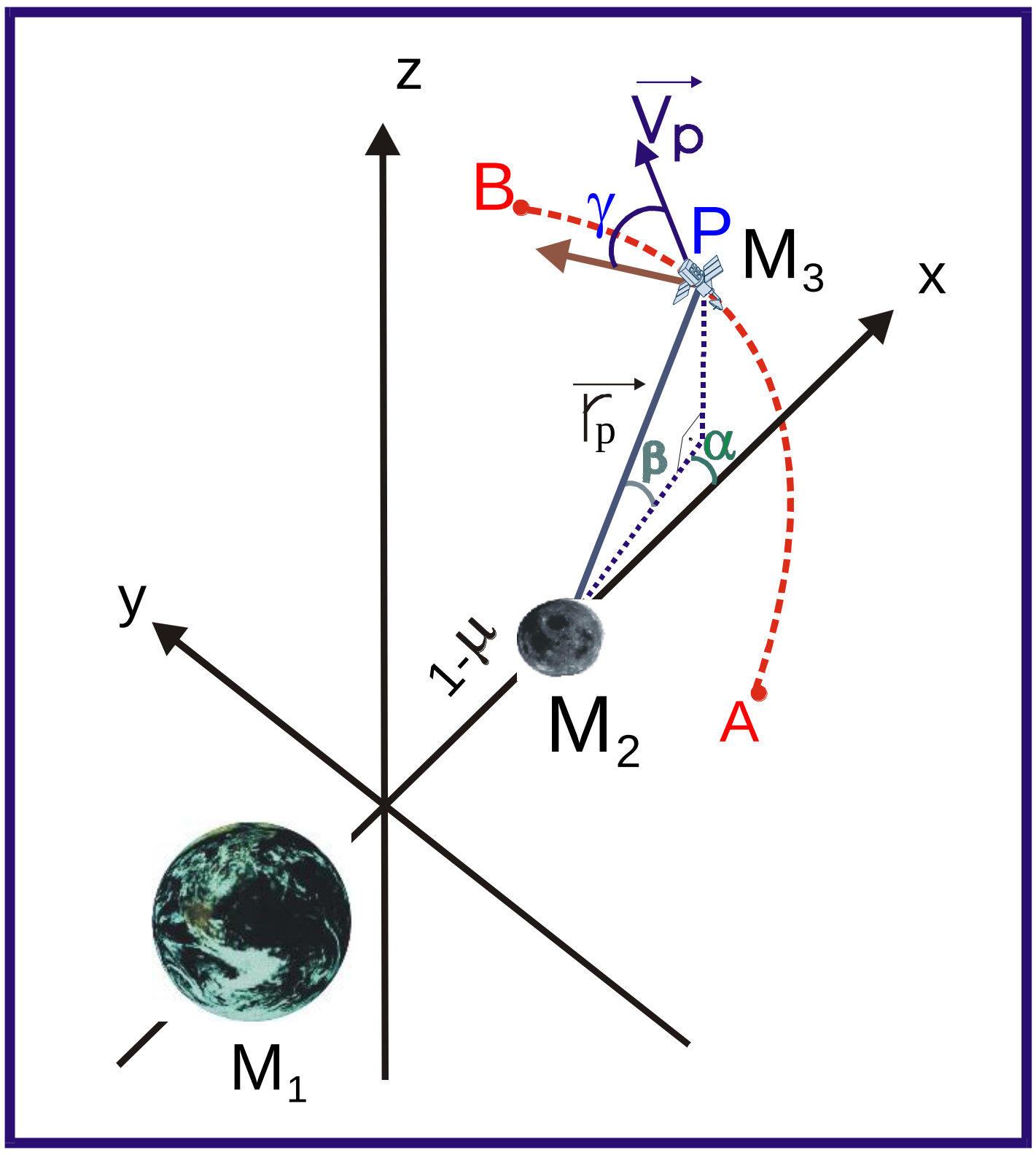

Fig. 1 - The Swing-By in Three Dimensions.

ANALYTICAL EQUATIONS FOR THE SWING-BY IN THREE DIMENSIONS

First, it is calculated the initial conditions with respect to $M_2$ at the periapsis. They are (see Fig. 1):

Position:

$$x_i = r_p \cos\beta \cos\alpha \tag{1}$$

$$y_i = r_p \cos\beta \sin\alpha \tag{2}$$

$$z_i = r_p \sin\beta \tag{3}$$

Velocity:

$$V_{xi} = -V_p \sin\gamma \sin\beta \cos\alpha - V_p \cos\gamma \sin\alpha \tag{4}$$

$$V_{yi} = -V_p \sin\gamma \sin\beta \sin\alpha + V_p \cos\gamma \cos\alpha \tag{5}$$

$$V_{zi} = V_p \cos\beta \sin\gamma \tag{6}$$

During the passage, it is assumed that the two-body celestial mechanics are valid and the whole maneuver takes place in the plane defined by the vectors $\vec{r}_p$ and $\vec{v}_p$. So, the vectors $\vec{v}_\infty^-$ and $\vec{v}_\infty^+$, that are velocity vectors before and after the swing-by, respectively, with respect to $M_2$ can be written as a linear combination of the versors associated with $\vec{r}_p$ and $\vec{v}_p$. Using $\vec{v}_\infty$ to represent both $\vec{v}_\infty^-$ and $\vec{v}_\infty^+$, since the conditions are the same for both vectors and a double solution will give the values for $\vec{v}_\infty^-$ and $\vec{v}_\infty^+$, we have:

$$\vec{V}_\infty = A\frac{\vec{r}_p}{r_p} + B\frac{\vec{V}_p}{V_p} \tag{7}$$

Which means that:

$$\begin{aligned}\vec{V}_\infty = {} & A(\cos\beta\cos\alpha, \cos\beta\sin\alpha, \sin\beta) + \\ & + B(-\sin\gamma\sin\beta\cos\alpha - \cos\gamma\sin\alpha, \sin\gamma\sin\beta\sin\alpha + \\ & + \cos\gamma\cos\alpha, \cos\beta\sin\gamma)\end{aligned} \tag{8}$$

With A, B constants that follows the relations:

$A^2 + B^2 = V_\infty^2$, where $v_\infty$ can be obtained from $V_\infty^2 = V_p^2 - \frac{2\mu}{r_p}$, that represents the conservation of energy of the two-body dynamics. A second requirement for $\vec{v}_\infty$ is that it makes an angle $\delta$ with $\vec{v}_p$, where $\delta$ is half of the total rotation angle described by the velocity vector during the maneuver (angle between $\vec{v}_\infty^-$ and $\vec{V}_\infty^+$). This condition can be written as:

$$\vec{V}_\infty \bullet \vec{V}_p = V_\infty V_p \cos\delta \tag{9}$$

where the dot represents the scalar product between two vectors.

From the two-body celestial mechanics it is known that:

$$\sin\delta = \frac{1}{1 + \frac{r_p V_\infty^2}{\mu_2}} \tag{10}$$

Using the equation for $\vec{v}_\infty$ as a function of $\vec{r}_p$ and $\vec{v}_p$, we have:

$$\vec{V}_\infty \bullet \vec{V}_p = \left( A\frac{\vec{r}_p}{r_p} + B\frac{\vec{V}_p}{V_p} \right) \bullet \vec{V}_p = BV_p = V_\infty V_p \cos\delta \tag{11}$$

So, $B = V_\infty \cos\delta$, because $\vec{r}_p \bullet \vec{v}_p = 0$ (at the periapsis $\vec{r}_p$ and $\vec{v}_p$ are perpendicular) and $\vec{v}_p \bullet \vec{v}_p = v_p^2$.

Then, since $A^2 + B^2 = v_\infty^2 \Rightarrow A^2 = v_\infty^2 - B^2 = v_\infty^2 - v_\infty^2 \cos^2\delta = v_\infty^2(1-\cos^2\delta) = v_\infty^2 \sin^2\delta \Rightarrow A = \pm V_\infty \sin\delta$

From those conditions, we have:

$$\begin{aligned}\vec{V}_\infty^- = {} & V_\infty \sin\delta(\cos\beta\cos\alpha, \cos\beta\sin\alpha, \sin\beta) + \\ & + V_\infty \cos\delta(-\sin\gamma\sin\beta\cos\alpha - \cos\gamma\sin\alpha, \\ & - \sin\gamma\sin\beta\sin\alpha + \cos\gamma\cos\alpha, \cos\beta\sin\gamma)\end{aligned} \tag{12}$$

$$\begin{aligned}\vec{V}_\infty^+ = {} & -V_\infty \sin\delta(\cos\beta\cos\alpha, \cos\beta\sin\alpha, \sin\beta) + \\ & + V_\infty \cos\delta(-\sin\gamma\sin\beta\cos\alpha - \cos\gamma\sin\alpha, \\ & - \sin\gamma\sin\beta\sin\alpha + \cos\gamma\cos\alpha, \cos\beta\sin\gamma)\end{aligned} \tag{13}$$

For $M_2$, its velocity with respect to an inertial frame ($\vec{v}_2$) is assumed to be:

$$\vec{V}_2 = (0, V_2, 0) \tag{14}$$

By using vector addition:

$$\begin{aligned}\vec{V}_i = \vec{V}_\infty^- + \vec{V}_2 = {} & V_\infty \sin\delta(\cos\beta\cos\alpha, \cos\beta\sin\alpha, \sin\beta) + \\ & + V_\infty \cos\delta(-\sin\gamma\sin\beta\cos\alpha - \cos\gamma\sin\alpha, -\sin\gamma\sin\beta\sin\alpha + \\ & + \cos\gamma\cos\alpha, \cos\beta\sin\gamma) + (0, V_2, 0)\end{aligned} \tag{15}$$

$$\vec{V}_0 = \vec{V}_\infty^+ + \vec{V}_2 = -V_\infty \sin\delta(\cos\beta\cos\alpha, \cos\beta\sin\alpha, \sin\beta) +$$
$$+ V_\infty \cos\delta(-\sin\gamma\sin\beta\cos\alpha - \cos\gamma\sin\alpha, -\sin\gamma\sin\beta\sin\alpha +$$
$$+ \cos\gamma\cos\alpha, \cos\beta\sin\gamma) + (0, V_2, 0) \qquad (16)$$

where $\vec{v}_i$ and $\vec{v}_0$ are the velocity of the spacecraft with respect to the inertial frame before and after the swing-by, respectively.

From those equations, it is possible to obtain expressions for the variations in velocity, energy and angular momentum. They are:

$$\Delta\vec{V} = \vec{V}_0 - \vec{V}_i = -2V_\infty \sin\delta(\cos\alpha\cos\beta, \cos\beta\sin\alpha, \sin\beta) \qquad (17)$$

which implies that:

$$\Delta V = \left|\Delta\vec{V}\right| = 2V_\infty \sin\delta \qquad (18)$$

$$\Delta E = \frac{1}{2}\left(V_0^2 - V_i^2\right) = -2V_2 V_\infty \cos\beta\sin\alpha\sin\delta \qquad (19)$$

For the angular momentum ($\vec{C}$) the results are:

$$\vec{C}_i = \vec{R} \times \vec{V}_i = d\,V_\infty(0, -\sin\beta\sin\delta + \cos\beta\cos\delta\sin\gamma,$$
$$\frac{V_2}{V_\infty} + \cos\alpha\cos\delta\cos\gamma + \cos\beta\sin\alpha\sin\delta - \cos\delta\sin\alpha\sin\beta\sin\gamma) \qquad (20)$$

$$\vec{C}_0 = \vec{R} \times \vec{V}_0 = d\,V_\infty(0, \sin\beta\sin\delta - \cos\beta\cos\delta\sin\gamma,$$
$$\frac{V_2}{V_\infty} + \cos\alpha\cos\delta\cos\gamma - \cos\beta\sin\alpha\sin\delta - \cos\delta\sin\alpha\sin\beta\sin\gamma) \qquad (21)$$

Where $\vec{R} = (d, 0, 0)$ is the position vector of $M_2$.

Then:

$$\Delta\vec{C} = \vec{C}_0 - \vec{C}_i = 2d\,V_\infty \sin\delta(0, \sin\beta, -\cos\beta\sin\alpha) \qquad (22)$$

and $\left|\Delta\vec{C}\right| = 2dV_\infty \sin\delta\left(\cos^2\beta\sin^2\alpha + \sin^2\beta\right)^{1/2}$ (23)

Using the definition of angular velocity $\omega = \frac{V_2}{d}$ it is possible to get:

$$\omega\Delta C_Z = -2V_2V_\infty\cos\beta\sin\alpha\sin\delta = \Delta E \quad (24)$$

For the inclination, the results are the following:

---

$$\left|\vec{C}_i\right| = dV_\infty\left(\left(\sin\beta\sin\delta + \cos\beta\cos\delta\sin\gamma\right)^2 + \left(\frac{V_2}{V_\infty} + \cos\alpha\cos\delta\cos\gamma + \cos\beta\sin\alpha\sin\delta - \cos\delta\sin\alpha\sin\beta\sin\gamma\right)^2\right)^{1/2}$$

(25)

$$C_{iZ} = dV_\infty\left(\frac{V_2}{V_\infty} + \cos\alpha\cos\delta\cos\gamma + \cos\beta\sin\alpha\sin\delta - \cos\delta\sin\alpha\sin\beta\sin\gamma\right) \quad (26)$$

So, $$\operatorname{Cos}(i_i) = \frac{C_{iZ}}{\left|\vec{C}_i\right|} = \frac{1}{\sqrt{1+\left(\dfrac{\sin\beta\sin\delta + \cos\beta\cos\delta\sin\gamma}{\dfrac{V_2}{V_\infty} + \cos\alpha\cos\delta\cos\gamma + \cos\beta\sin\alpha\sin\delta - \cos\delta\sin\alpha\sin\beta\sin\gamma}\right)^2}} \quad (27)$$

$$\left|\vec{C}_o\right| = dV_\infty\left(\left(\sin\beta\sin\delta - \cos\beta\cos\delta\sin\gamma\right)^2 + \left(\frac{V_2}{V_\infty} + \cos\alpha\cos\delta\cos\gamma - \cos\beta\sin\alpha\sin\delta - \cos\delta\sin\alpha\sin\beta\sin\gamma\right)^2\right)^{1/2} \quad (28)$$

$$C_{oZ} = dV_\infty\left(\frac{V_2}{V_\infty} + \cos\alpha\cos\delta\cos\gamma - \cos\beta\sin\alpha\sin\delta - \cos\delta\sin\alpha\sin\beta\sin\gamma\right) \quad (29)$$

So,\

$$\operatorname{Cos}(i_o) = \frac{C_{oZ}}{\left|\vec{C}_o\right|} = \frac{1}{\sqrt{1+\left(\dfrac{\sin\beta\sin\delta - \cos\beta\cos\delta\sin\gamma}{\dfrac{V_2}{V_\infty} + \cos\alpha\cos\delta\cos\gamma - \cos\beta\sin\alpha\sin\delta - \cos\delta\sin\alpha\sin\beta\sin\gamma}\right)^2}} \quad (30)$$

---

Where $\vec{C}_i$ and $\vec{C}_o$ are the initial and final angular momentum, respectvely, $i_i$ and $i_o$ are the initial and final inclinations, respectevely, and the subscript Z stands for the z-component of the angular momentum.

---

The variation in inclination $\Delta i$ can be obtained from $i_0$-$i_i$.

For the planar maneuver ($\beta = \gamma = 0º$), those equations are reduced to the well-known results[18]:

$$\Delta E = -2V_2V_\infty \sin\alpha \sin\delta \tag{31}$$

$$\Delta V = 2V_\infty \sin\delta \tag{32}$$

$$\Delta C = 2dV_\infty \sin\alpha \sin\delta \tag{33}$$

The equations developed here show that:

i) There is an important result: $\Delta E = \omega \Delta C_Z$, that for the planar case can be simplified to $\Delta E = \omega \Delta C$, since the total variation of the angular momentum is in the z-direction;

ii) The variation in energy is the one obtained for the planar case multiplied by the factor cos($\beta$). So, the maximum variation occurs for the planar maneuver ($\beta$ = 0°) and the minimum, that is zero, for the polar passages ($\beta$ = 90°);

iii) The parameters $V_2$ and $V_\infty$ are positive quantities (they are the magnitude of two vectors), as well as sin($\delta$) (because $0 < \delta < 90°$) and cos($\beta$) (because $-90° < \beta < 90°$). Then, the only parameter that affects the sign of $\Delta E$ is sin($\alpha$). The conclusion is that for values of $\alpha$ in the range $0° < \alpha < 180°$, $\Delta E$ is negative (decrease in energy) and for $\alpha$ in the range $180° < \alpha < 360°$, $\Delta E$ is positive (increase in energy). So the final conclusions are: if the swing-by is in front of $M_2$ there is a decrease in the energy of $M_3$, with a maximum loss at $\alpha$ = 90° ($\Delta\vec{V}$ opposite to $\vec{V}_2$); if the swing-by is behind $M_2$ there is an increase in the energy of $M_3$, with a maximum gain at $\alpha$ = 270° ($\Delta\vec{V}$ aligned with $\vec{V}_2$);

iv) For the variation of the magnitude of the angular momentum the minima, with value zero, are located at $\beta$ = 0° and $\alpha$ = 0°, 180°, 360°. From those points the magnitude increases with the distance from the points. Fig. 2 shows this results for the system Sun-Jupiter in the case $r_p$ = 0.000137595 and $V_p$ = 4.0. The range used for $\alpha$ is $180° \le \alpha \le 360°$, because the interval $0° \le \alpha \le 180°$ is symmetric;

v) For the variation in the components of the angular momentum, it is possible to conclude that the x-component is always zero; the z-component is related to the variation in energy as shown before and it has a variation according to cos($\beta$), which implies that the minima are at the poles and the maximum occurs for a planar maneuver; the y-component has a variation according to sin(β), which implies that the maximum is at β = 90° and the minimum is at β = -90°, with value zero for a planar maneuver;
vi) The variation in velocity is independent of the angle β, so the equations for the planar maneuver are still valid;

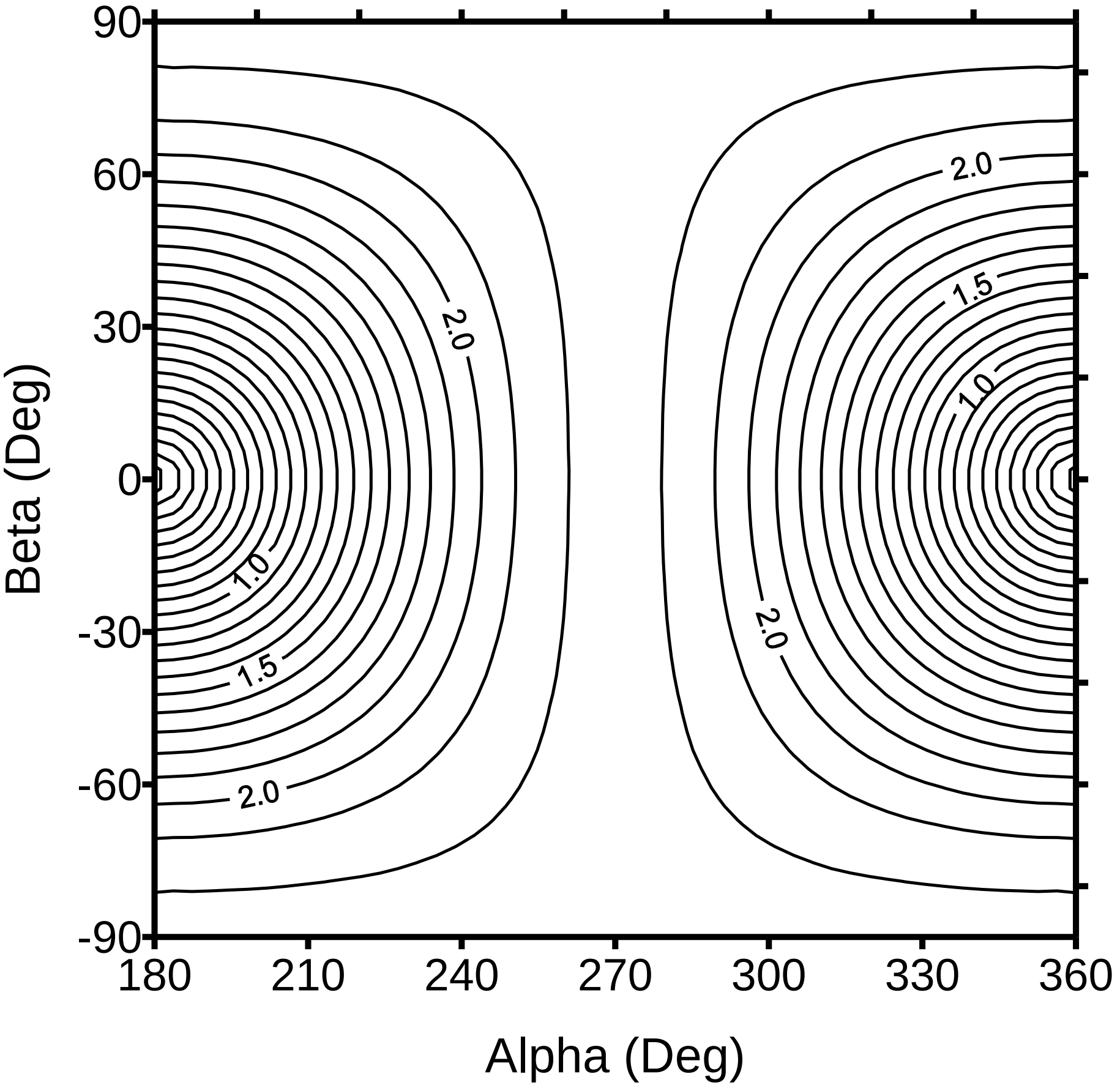

Fig. 2 - Variation in angular momentum for $r_p$ = 0.000137595 and $V_p$ = 4.0 (Sun-Jupiter system).

## ANALYSIS OF THE INCLINATION

An interesting question that appears in this problem is what happens to the inclination of the spacecraft due to the close approach. To investigate this fact, the equations that calculate the inclination of the trajectories before and after the closest approach are studied in more detail. Fig. 3-10 show results for the variation of the inclination for a series of initial conditions, considering the case □ = 0° and for the Sun-Jupiter system. This constraint is assumed, because it is the most usual situation in interplanetary research, since the planets have orbits that are almost coplanar. The horizontal axis represents the angle α, and the vertical axis represents the angle β. The variation in inclination is shown in the contour plots. All the angles are expressed in degrees and the velocities in canonical units (one canonical unit of velocity is the velocity of a spacecraft in a circular orbit with unitary radius). Those results are very similar to the numerical ones obtained by Prado[23].

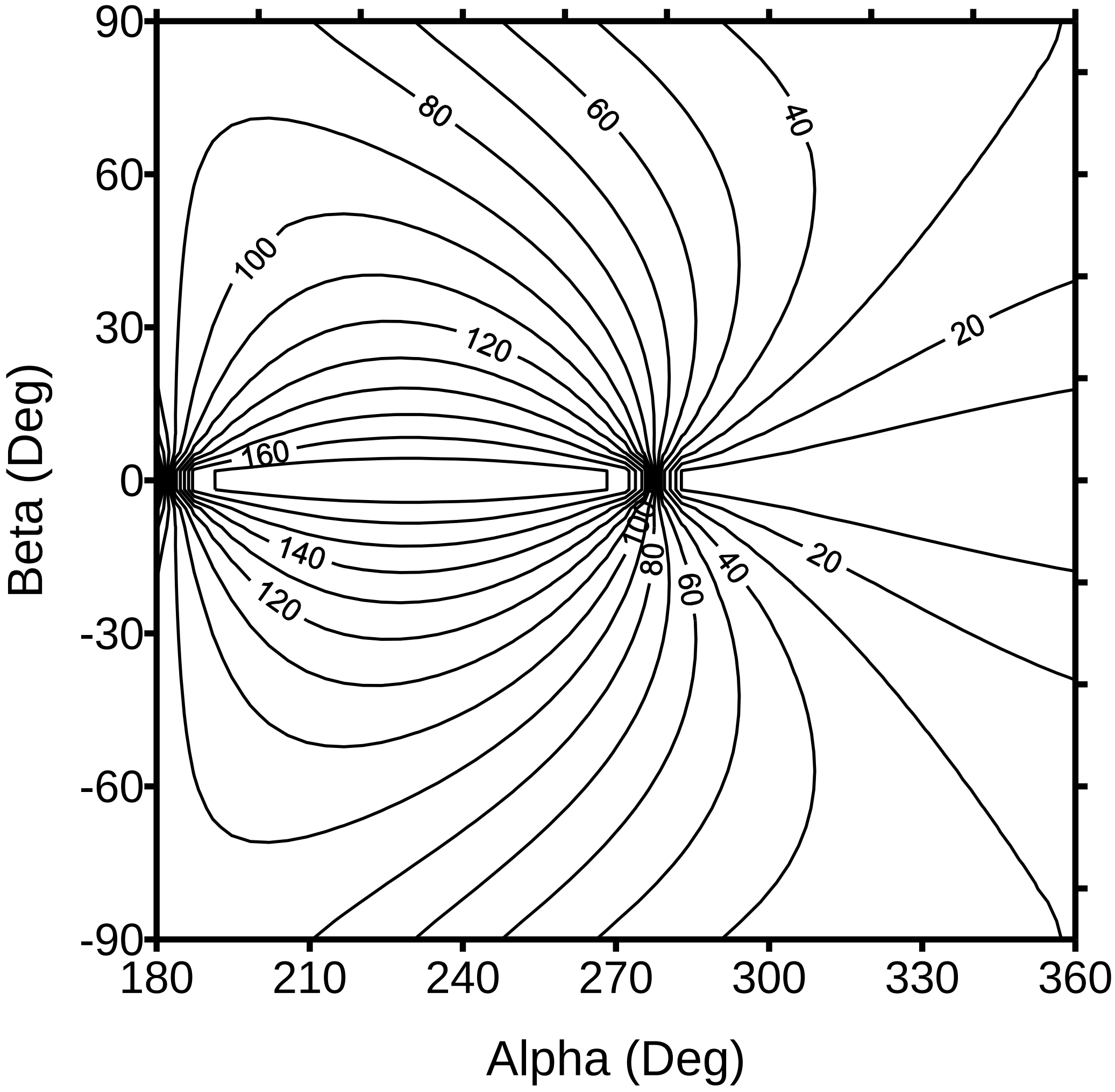

Fig. 3 - Inclination before swing-by for $r_p$ = 0.000137595, $V_p$ = 4.0.

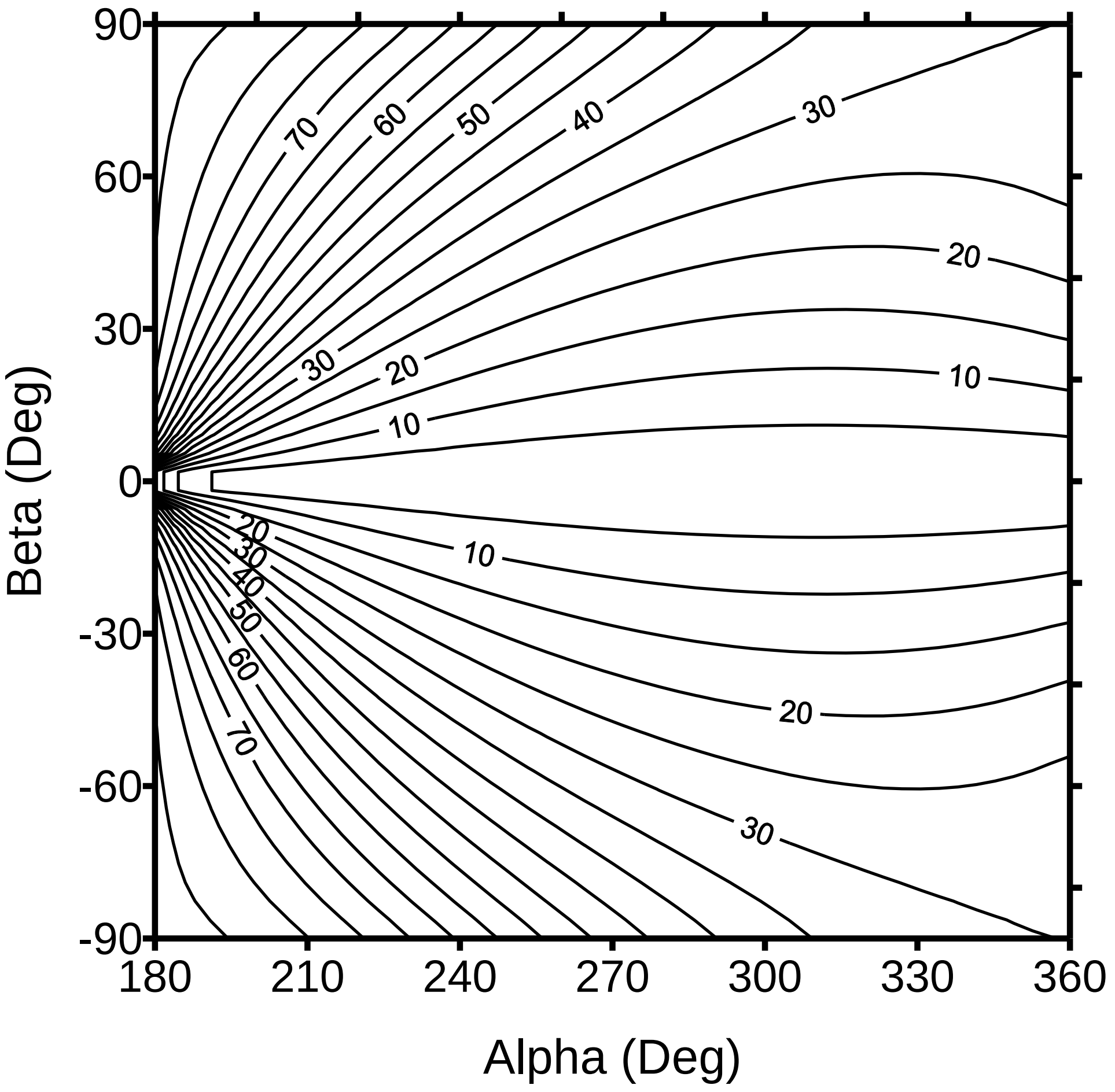

Fig. 4 - Inclination after swing-by for $r_p = 0.000137595$, $V_p = 4.0$.

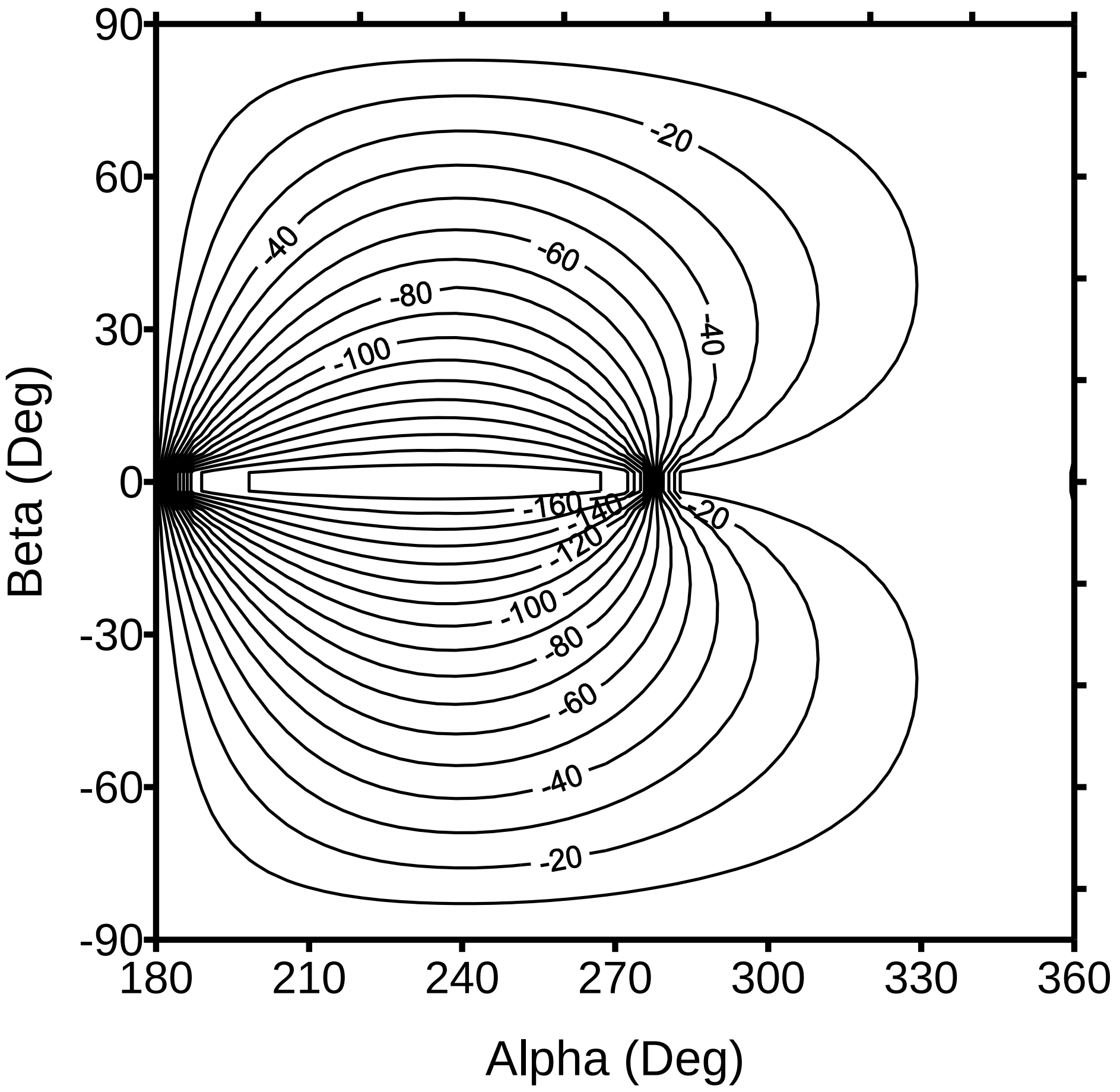

Fig. 5 - Variation in Inclination for $r_p$ = 0.000137595 and $V_p$ = 4.0.

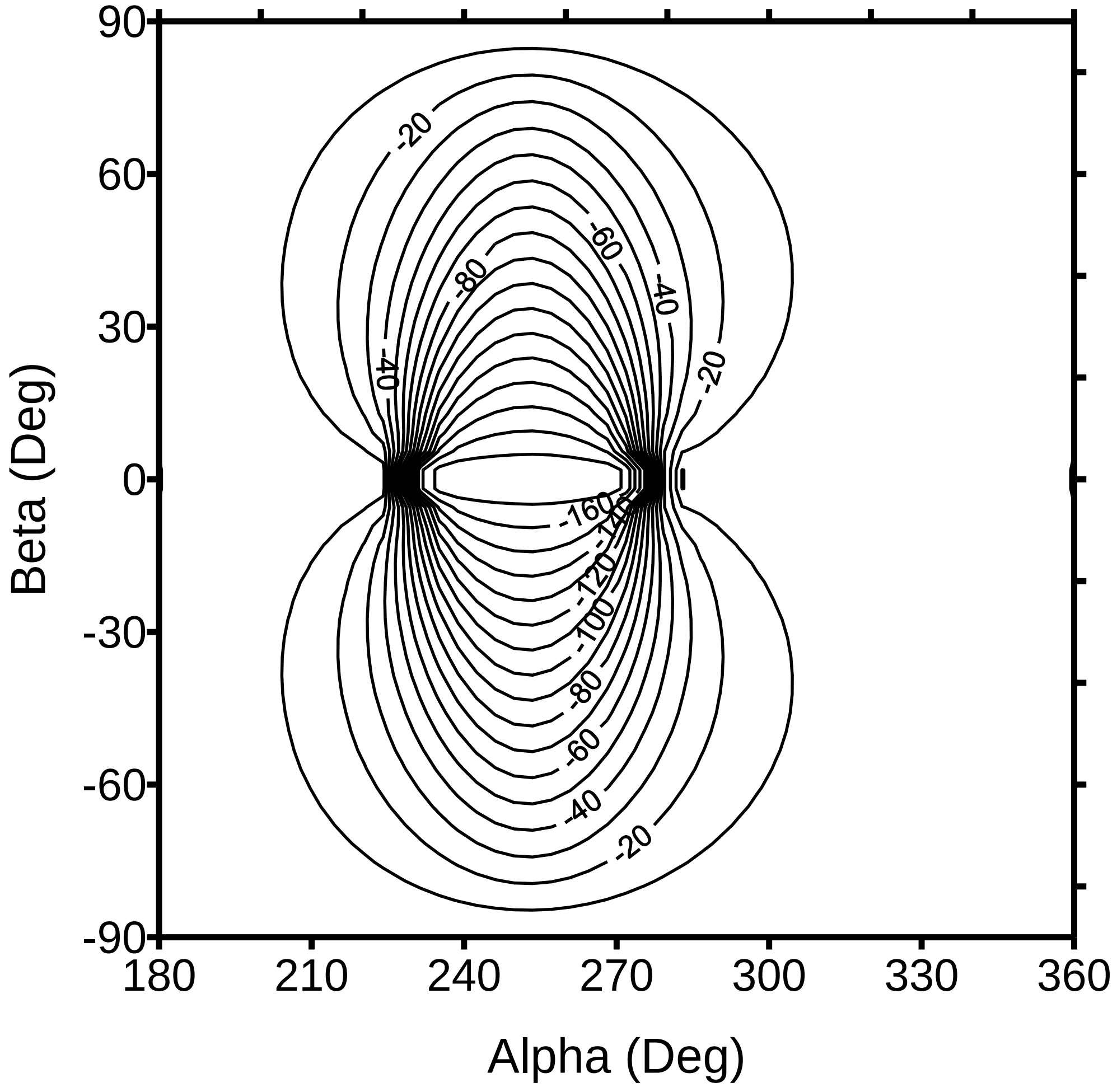

Fig. 6 - Variation in Inclination for $r_p$ = 0.000137595 and $V_p$ = 5.0.

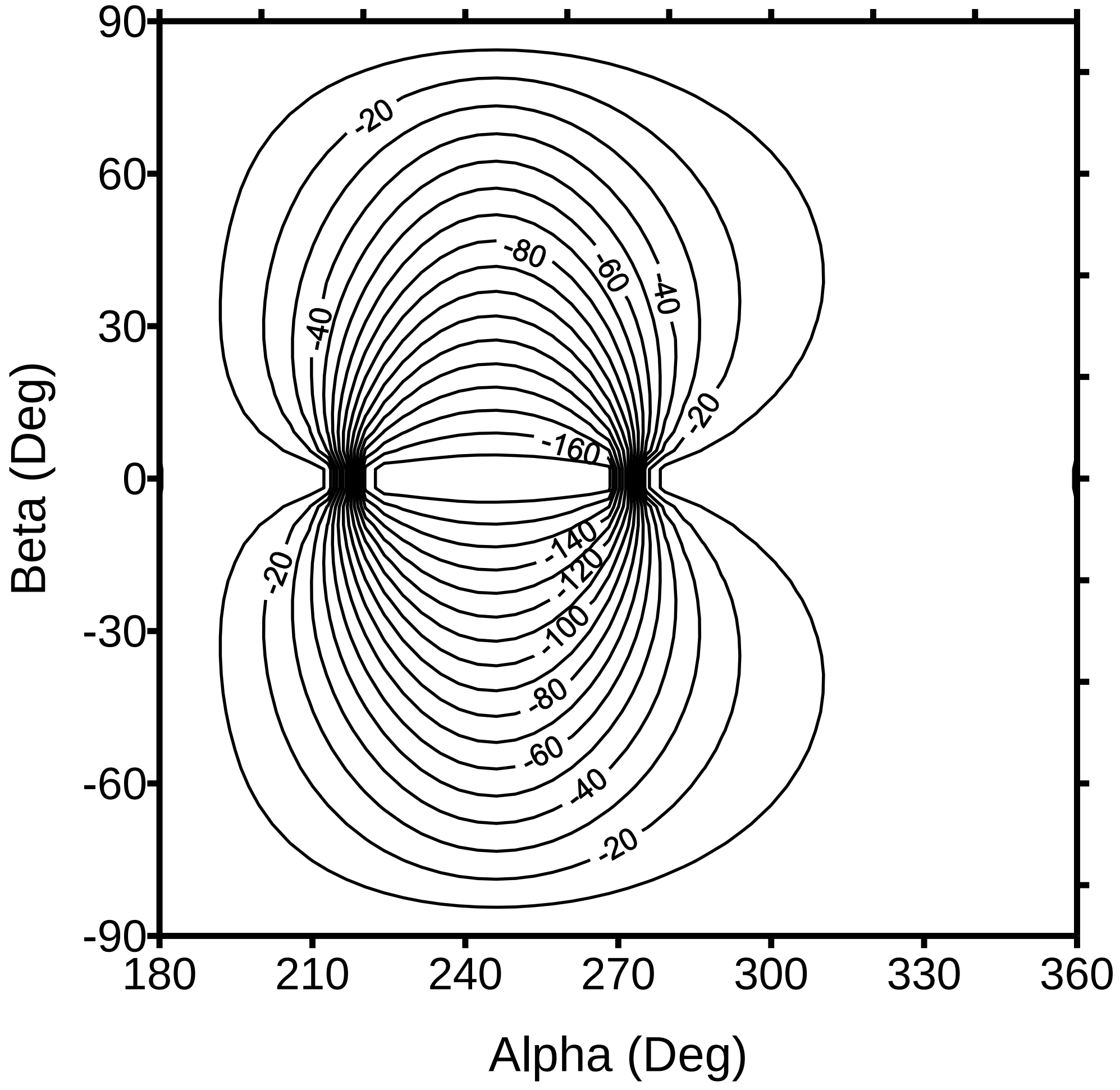

Fig. 7 - Variation in Inclination for $r_p$ = 0.000183460 and $V_p$ = 4.0.

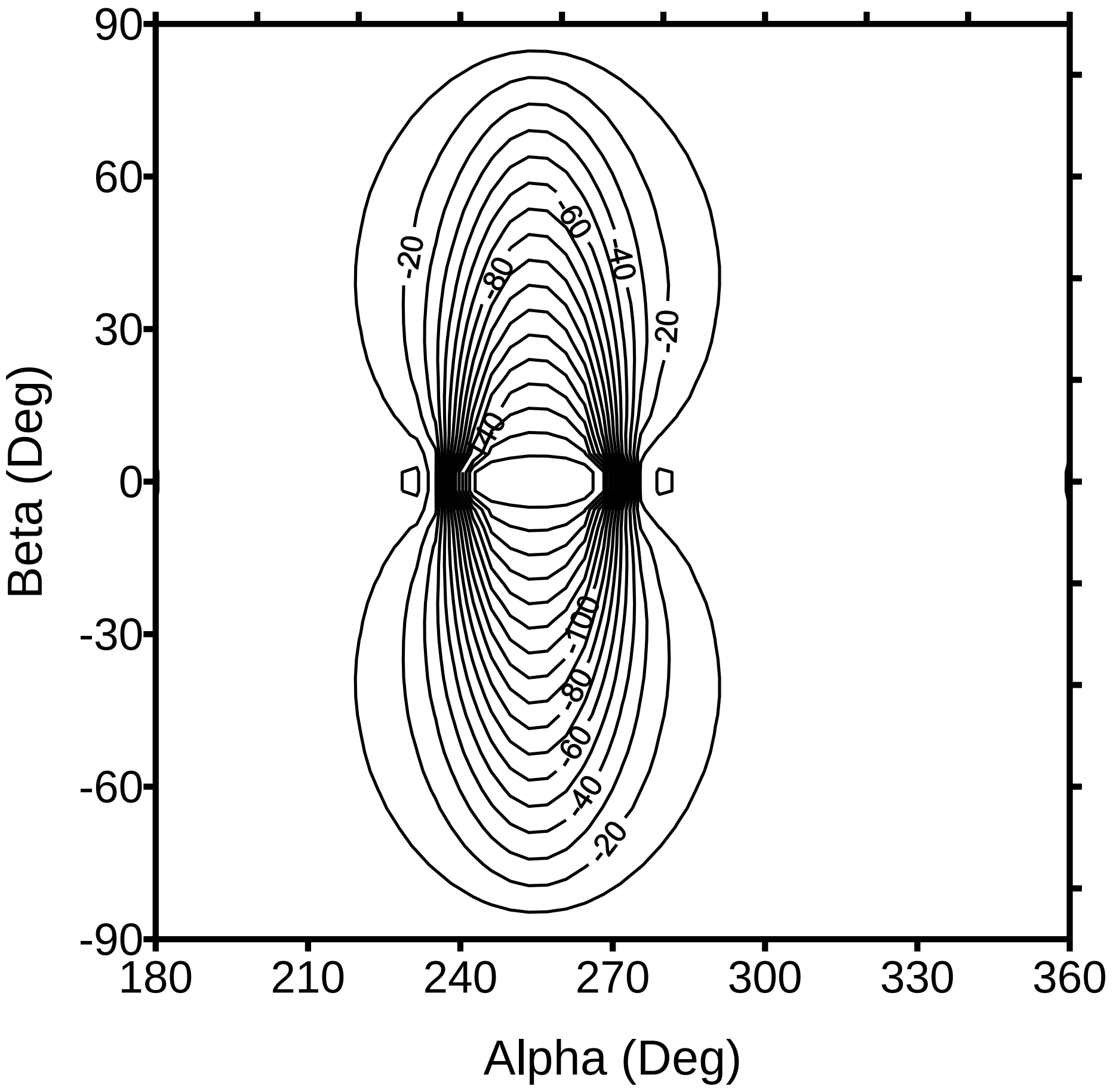

Fig. 8 - Variation in Inclination for $r_p$ = 0.000183460 and $V_p$ = 5.0.

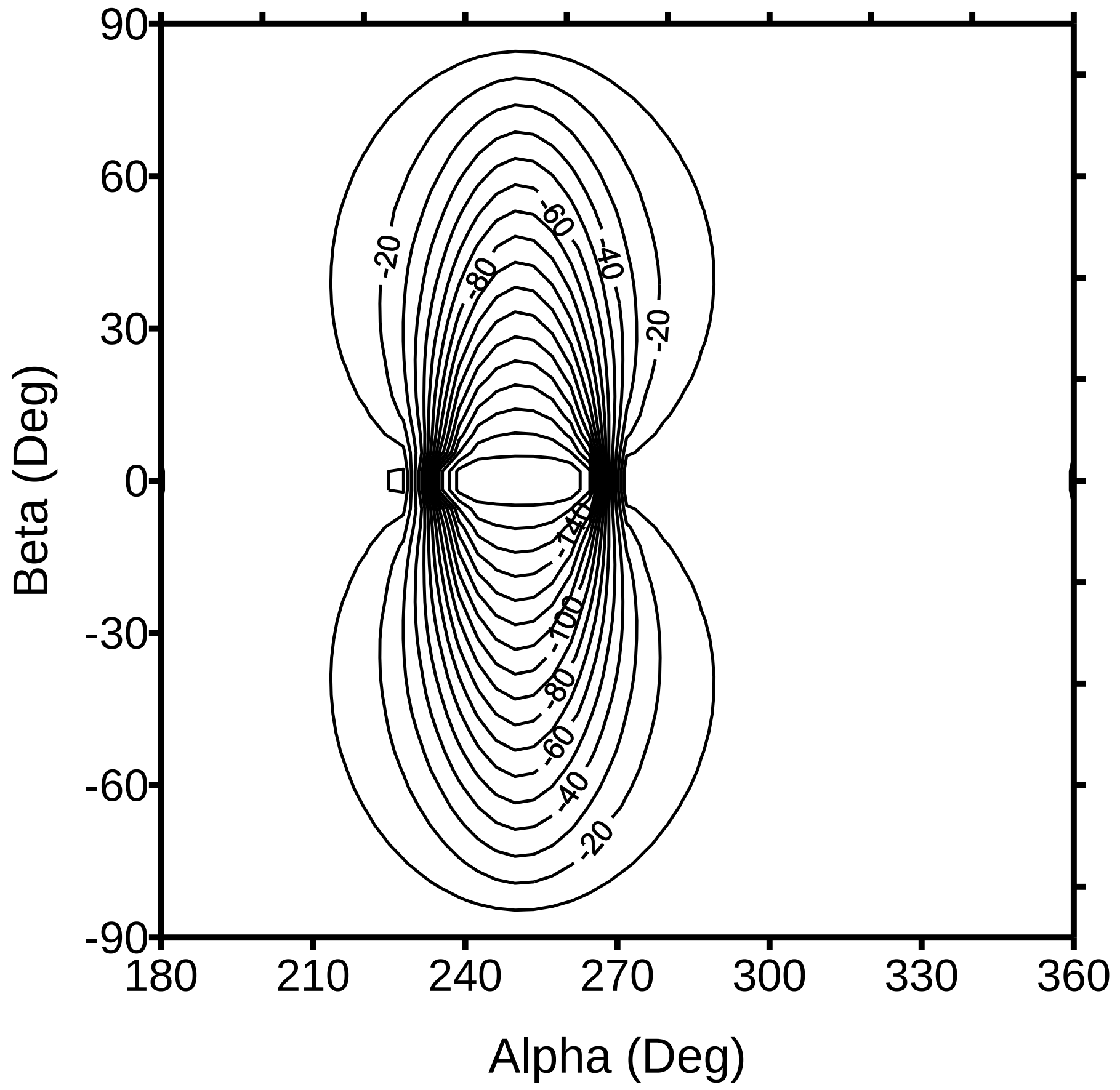

Fig. 9 - Variation in Inclination for $r_p$ = 0.000275190 and $V_p$ = 4.0.

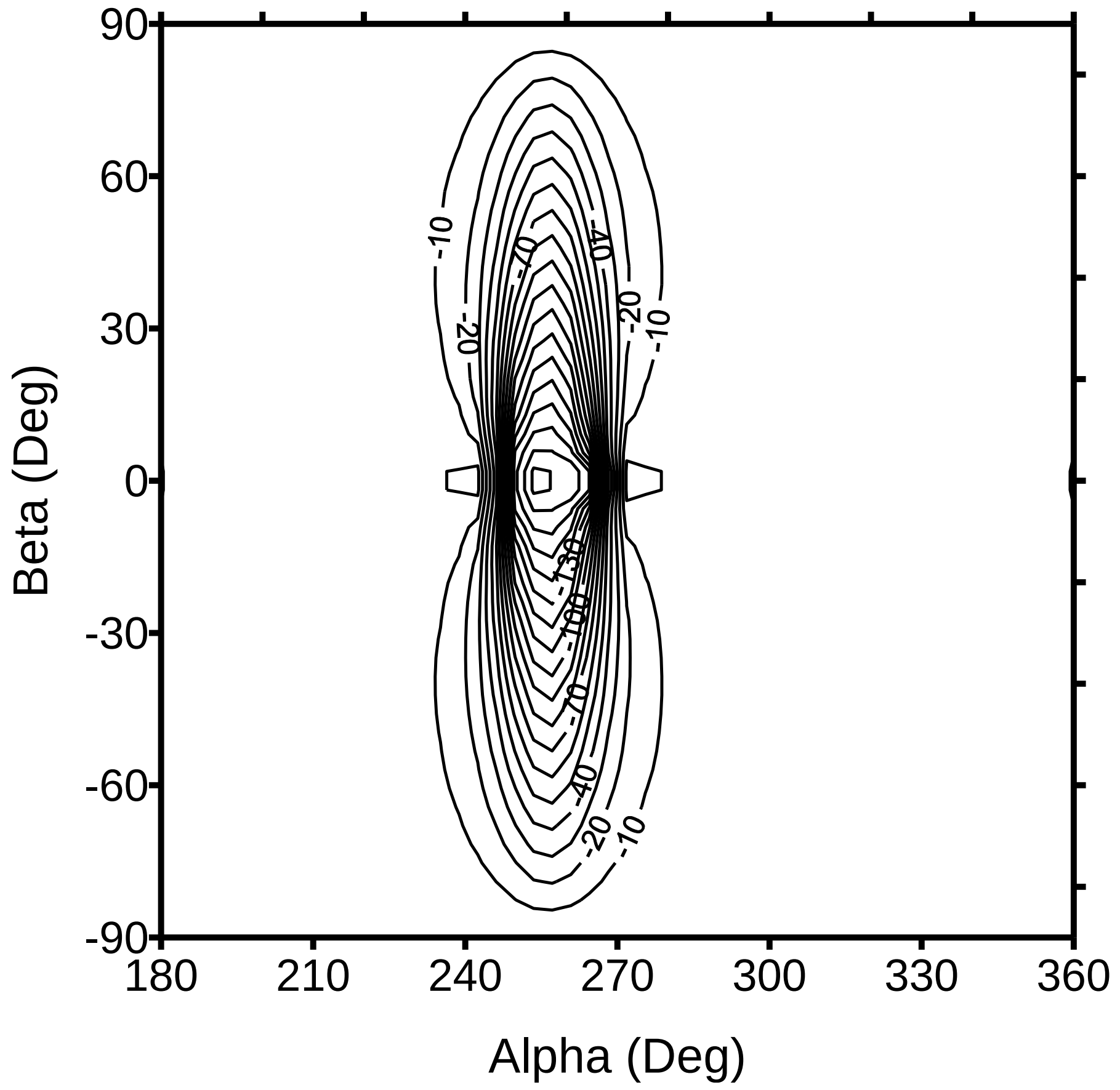

Fig. 10 - Variation in Inclination for $r_p$ = 0.000275190 and $V_p$ = 5.0.

The interval of $\alpha$ is 180° < $\alpha$ < 360° because there is a symmetry in the system and the values for the variation in inclination in the interval 0° < $\alpha$ < 180° are the same ones for the interval 180° < $\alpha$ < 360° with a reversal of sign. So, positive values for the variation in inclination are in the symmetric part of the plots (not shown here) and negative values are in the regions shown. Several conclusions come from those results. The most interesting ones are: i) when $\beta$ = 0º (planar maneuver) the variation in inclination can have only three possible values: $\pm 180^o$, for a maneuver that reverse the sense of its motion, or 0º for a maneuver that does not reverse its motion. Those results agree with the physical-model, since the fact that $\beta$ = 0º implies in a planar maneuver that does not allow values for the inclination other than 0º or 180º. This is clearly shown in the figures, following the line $\beta$ = 0º. The plots are divided in two parts: one with $\Delta i = \pm 180^o$ and the other one with $\Delta i = 0^o$; ii) Looking at any vertical line (a line of constant $\alpha$) it is clear that the change in inclination goes to zero at the poles $(\beta = \pm 90^o)$. This fact can be seen in the analytical equations because the difference in the equations for the inclination before and after the swing-by is a reversal in the sign of the terms that are multiplied by cos($\beta$). So, if this term is null, there is no variation in the inclination.

It is also clear that the variation in inclination is symmetric with respect to the angle $\beta$ ($+\beta$ and $-\beta$ generate the same $\Delta i$); iii) when $\alpha = 0º$ or $\alpha = 180º$ there is no change in the inclination. This is in agreement with the fact that a maneuver with this geometry does not change the trajectory at all. Looking at any horizontal line (a line of constant $\beta$) it is visible that this curve has a maximum in the magnitude of $\Delta i$ somewhere between the two fixed zeroes at $\alpha = 0º$ and $\alpha = 180º$. This result is valid only in the situation $\gamma = 0°$. This maximum can be calculated by the analytic equations, if necessary; iv) when the periapsis distance or the velocity at periapsis increases, the effects of the swing-by in the maneuver are reduced. In the plots shown, this can be verified by the fact that the area of the regions where the variation in inclination is close to zero increases. This is the reason why the regions full of lines are reduced in the figures; v) The same is true when the velocity at periapsis increases.

Earth-Moon System

To generalize the results a little bit two simulations for the Earth-moon system are made. Figs. 11 and 12 show the results. It is clear that the same behavior of the Sun-Jupiter system is repeated. Those results are also very similar to the ones obtained by numerical integration by Felipe and Prado[24].

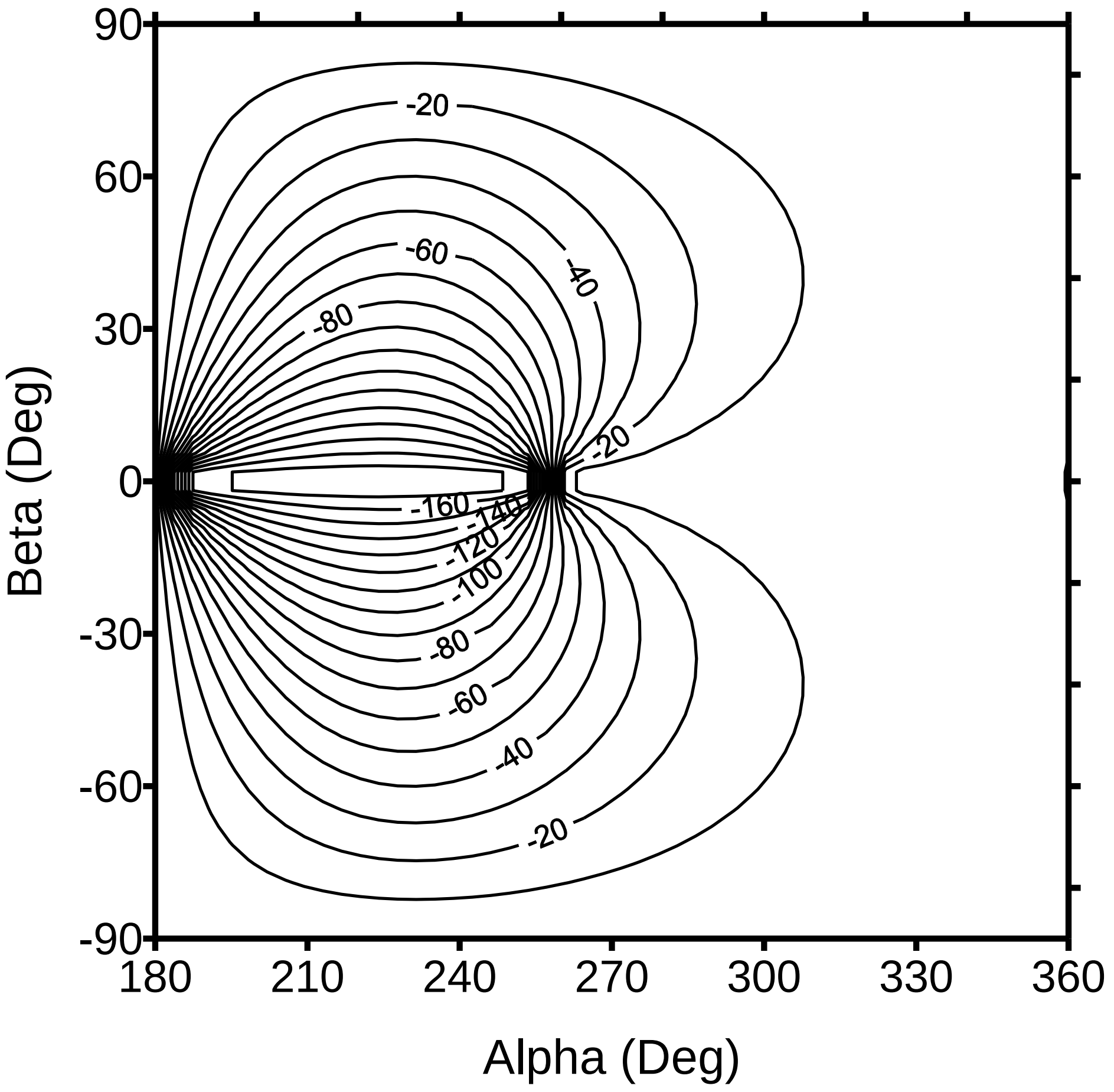

Fig. 11 - Variation in Inclination for the Earth-moon system with $r_p = 0.00476$ and $V_p = 2.6$.

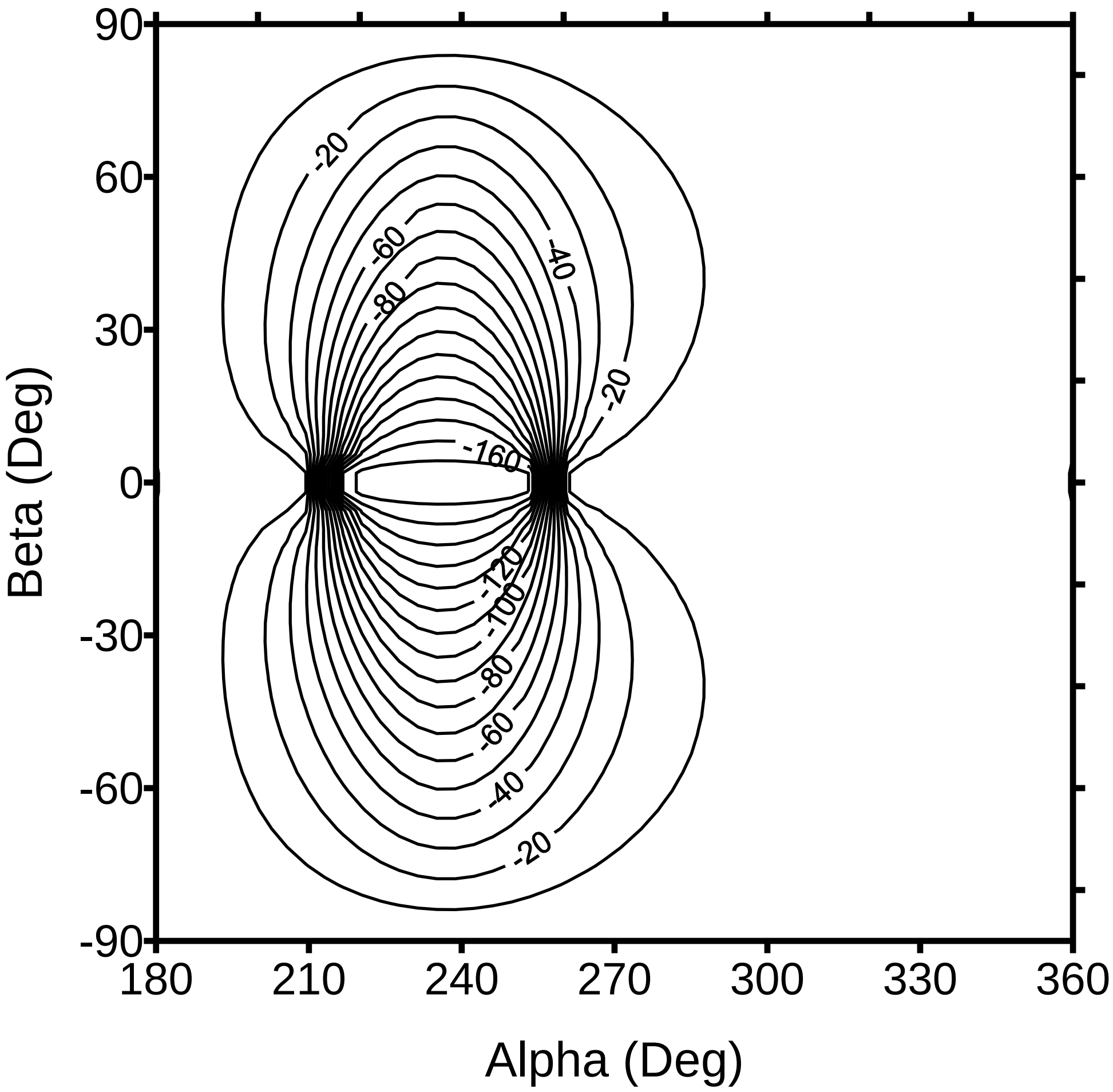

Fig. 12 - Variation in Inclination for the Earth-moon system with $r_p$ = 0.00675 and $V_p$ = 2.6.

<u>Applications</u>

An interesting application of the analytical equations for the inclinations before and after the passage developed here is to solve problems with given inclinations. The equations show that there are five independent parameters that determines the inclinations before and after the swing-by: $\alpha$, $\beta$, $\gamma$, $V_\infty$ (or the equivalent $V_p$) and $\delta$ (or the equivalent $r_p$, since $V_p$ is already given). Suppose that it is necessary to design a mission where the orbit before the close approach is fixed (so $\gamma$, $V_\infty$ and the inclination before the passage are given) and that it is desired to obtain a given inclination after the maneuver. The parameters $\alpha$, $\beta$ and $r_p$, can be obtained with small adjustments in the trajectory and they will be used as the control variables. Mathematically, there are five independent parameters to describe the maneuver, with two of them given ($\gamma$ and $V_\infty$) and two constraints between them (the inclinations before and after the passage are given), what leaves one free parameter. To solve the problem, the procedure used here is to specify a value for $r_p$ and solve the equations for the inclinations to find the variables $\alpha$, $\beta$. As an example, the following problem is used. It is desired to design a mission for a spacecraft to swing-by Jupiter with an orbit before the passage that implies in the following given parameters: $V_p = 4.5$, $\gamma = 0°$, inclination = 5°. It is required that the orbit after the swing-by to be polar, so the inclination is 90°. This means to maneuver the spacecraft to change an orbit close to planar to a perpendicular one, similar to the maneuver used for the Ulysses mission[25]. The results are shown in Fig. 13, that plots the periapsis distance (using the radius of Jupiter as the unit) in the horizontal axis and the angles $\alpha$ and $\beta$ in degrees in the vertical axis.

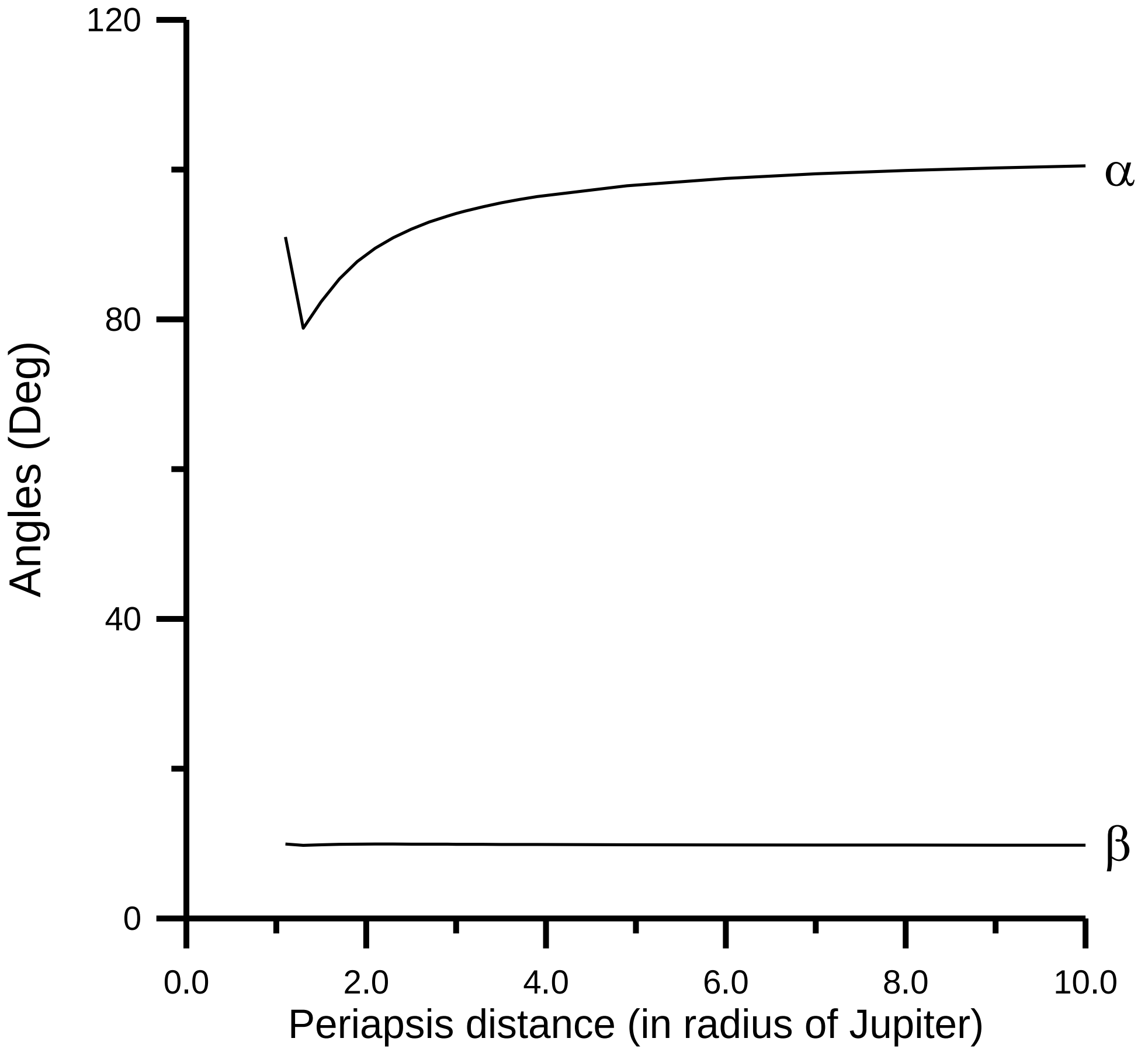

Fig. 13 – Solutions to maneuver a spacecraft between two orbits with given inclinations.

A slight different approach is to fix the initial orbit (so, the values of $\gamma$, $V_\infty$ and the inclination before the passage are given) and study the final inclination obtained as a function of $\alpha$ and $\beta$. In this approach, there are three independent parameters to choose: $\alpha$, $\beta$ and $r_p$ and one constraint: the initial inclination is given. The approach used here to solve this problem is to obtain the value of $r_p$ from the constraint equation and make a contour-plot of the inclination obtained after the passage as a function of the angles α (in the horizontal axis in degrees) and β (in the vertical axis in degrees). Fig. 14 show the solution for this situation, with the same initial conditions used in the previous example ($V_p$ = 4.5, γ = 0°, inclination before the passage = 5°). This figure explains the reason why the value obtained for β is almost constant. Looking in the centralpart of the plot (the region with final inclination close to 90°) it is clear that the lines of constant inclination after the passage are made by lines very close to horizontal, which means lines of constant β. The possibilities of multiple solutions are better observed, since there is a whole line of values that can solve the previous problem (given inclination after the passage).

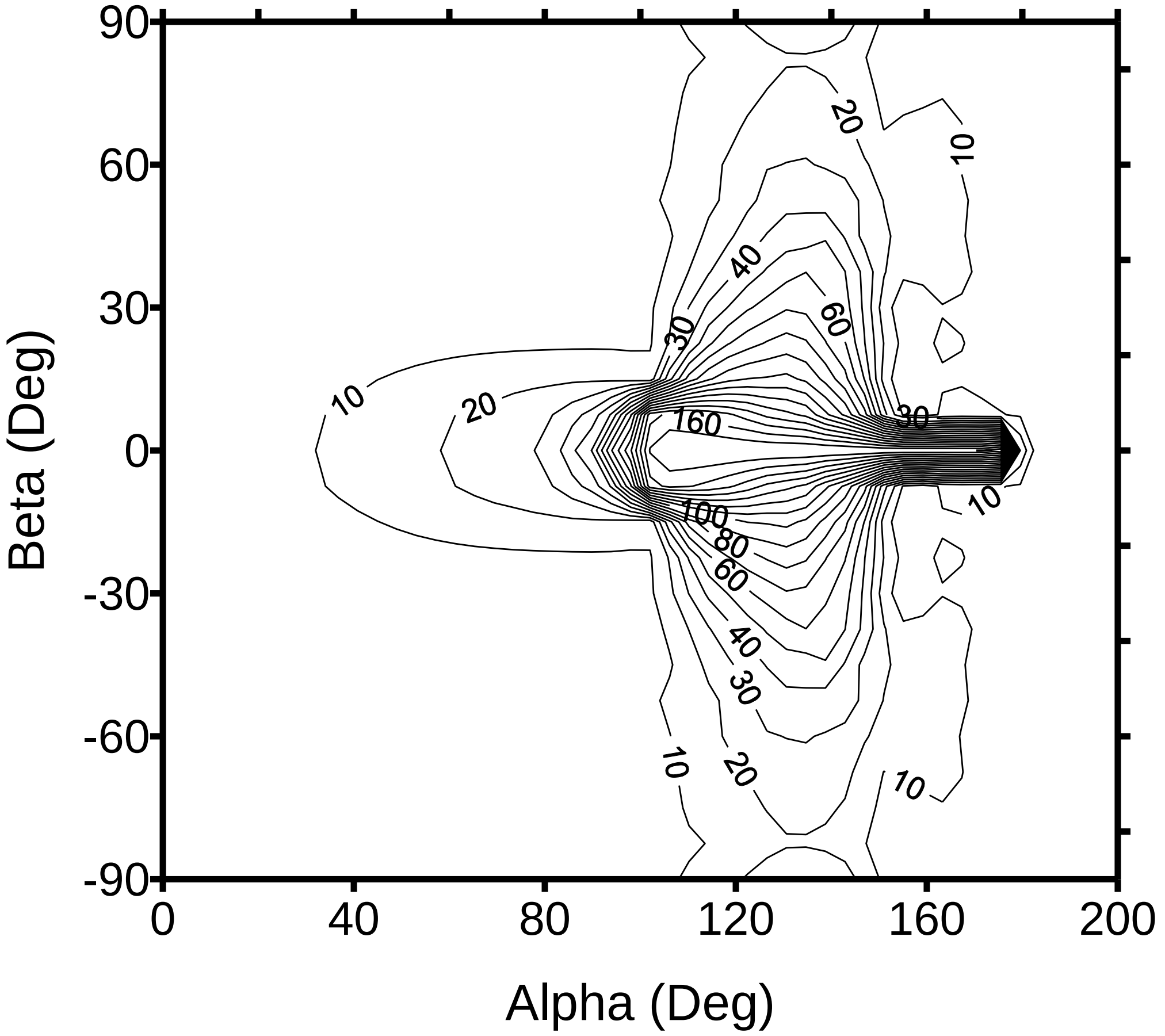

Fig. 14 – Inclinations obtained after the swing-by.

## CONCLUSIONS

In this paper, analytical equations based in the patched conics approximation were derived to calculate the variation in velocity, angular momentum, energy and inclination of a spacecraft that performs a swing-by maneuver. Several properties are derived and demonstrated. The most interesting ones are: i) for the planar maneuver the variation in inclination can have only three possible values: 180° and -180°, for a maneuver that reverse the sense of its motion, or 0º for maneuver that does not reverse its motion; ; ii) The change in inclination goes close to zero at the poles and it is symmetric with respect to the out of plane angle; iii) when $\alpha$, the angle between the periapsis line and the line connecting the two primaries is zero or 180º there is no change in the inclination. This result is valid only in the situation $\gamma = 0°$.

After that, the equations are used to solve practical problems in maneuvers that have fixed orbits before the swing-by, such as problems of obtained a given inclination after the swing-by.

## ACKNOWLEDGMENT

This paper has been supported by the Ministry of Science and Higher Education of the Russian Federation under Agreement No. FSSF-2024-0005.